\documentstyle[amssymb,epsfig,prb,aps]{revtex}

\begin{document}

\twocolumn[\hsize\textwidth\columnwidth\hsize\csname@twocolumnfalse\endcsname

\draft
\title{Transport spin polarization of
Ni$_{x}$Fe$_{1-x }$: electronic kinematics and
band structure.}
\author{B. Nadgorny$^{1}$, R. J. Soulen, Jr.$^{ 1}$, M. S. Osofsky$^{1}$, I. I.
Mazin$^{1}$, G. Laprade$^{2}$, R.J.M. van de Veerdonk$^{3,4}$, A.A. Smits$^4$,
S. F. Cheng$^{1}$, E. F. Skelton$^{1}$, S. B. Qadri$^{1}$}
\address{$^{1}$Naval Research Laboratory, Washington, DC 20375\\
$^{2}$INSA, Toulouse, France\\
$^3$Massachusets Institute of Technology, Cambridge, MA 02139\\
$^4$Eindhoven Institute of Technology, 5600 MB Eindhoven, The Netherlands}

\date{May 4, 1999}
\maketitle

\begin{abstract}
We present measurements of the transport spin polarization of
Ni$_x$Fe$_{1-x}\ (0 \leq
x \leq 1)$ using the recently-developed Point Contact Andreev
Reflection technique, and compare them with our
first principles calculations of the spin
polarization for this system. Surpisingly, the measured
spin polarization is almost composition-independent.
The results
clearly
demonstrate that the sign of the transport spin polarization does not coincide
with that of the difference of the densities of states at the Fermi
level.
Calculations
indicate that the independence of the spin polarization of the composition
is due to compensation of density of states and Fermi velocity in the 
s- and d- bands.
\end{abstract}
]

Spin-polarized transport in magnetic materials is beginning to play an
increasingly important role in  fundamental and applied research due to
the rapid advance of magnetoelectronics\cite{1}. The very definition
of this new field, based on the ability of magnetic metals to carry
spin-dependent current, implies that many physical phenomena and device
applications are determined by the
interplay of magnetic and transport
properties of these materials. Although many materials are
spin-polarized, technical constraints limit the number actually used in
practice to only a handful.
In particular, permalloy, a member of a family of binary alloys,
Ni$_{x}$Fe$ _{1-x}$ ($x=0.8$), features an attractive
combination of vanishingly small magnetostriction, low coercivity and high
permeability, which makes it the material of choice for magnetic recording
media, sensors, and nonvolatile magnetic random access memory.

Impressive progress in understanding 
magnetic properties of 3d-transition metal ferromagnets has
been made in the last decades, particularly due to
the advances of the band structure calculations,
based on the local spin
density approximation (LSDA). However,
  many aspects of the deceptively simple model
system of Ni$_{x}$Fe$_{1-x}$ alloys still elude quantitative
explanations.
One of the unresolved problems is the difficulty in
reconciling the itinerant character of magnetic d-electrons (which seems to be
reliably established by de-Haas-van Alfen experiments
  \cite{3}) and the value as well as the positive sign of the spin polarization
measured by tunneling experiments. Since the
electronic density of states (DOS) at the Fermi level is higher for spin-down
d-electrons than for s-electrons, it is obvious that this effect cannot 
be explained
within simple models based on the constant tunneling matrix element 
approximation.
The qualitative explanation was suggested in a number of papers (e.g.,Ref.
\onlinecite{4}), where it was pointed out that the tunneling matrix elements
for s- electrons are larger than for d- electrons.
Although this picture is instructive for a qualitative understanding
of the transport spin polarization, it is not very useful for
quantitative analsyis, since in transition metals electrons can be only
marginally divided into s- and d- types. Instead, it is more appropriate to
speak in terms of different bands with different Fermi
velocities. Within this approach we propose a natural quantitative
interpretation of this effect based on band structure calculations,
consistent with our spin polarization measurements and the most recent
tunneling results\cite {9}.

In order to make a meaningful comparison between spin polarization
measurements in various experiments and the theory, the spin polarization
must
be clearly defined\cite {6,M}. One cannot generally expect that the
tunneling spin polarization, $P_T$, which is determined by a fraction of the
spin-polarized current, is the same as the spin polarization probed, 
for instance,
by spin-resolved photoemission, $P_N$. While $P_N$ is related to the electronic
density of states (DOS) at the Fermi surface, $N(E_{F})$, $P_T$ is 
determined by
a weighted average of the DOS and tunneling matrix elements, which are,
in general, functions of the Fermi velocities.  In the spin-polarized Andreev
reflection experiments\cite{6,B}, yet another spin
polarization, $P_A$, is measured.
In the clean (ballistic or Sharvin) limit, $P_A$
is defined by the average projection of the Fermi velocity $%
v_{F}$ on {\it z}, the direction normal to the contact
plane, and thus $P_A=P_{Nv}\propto
  \left\langle N(E_{F})v_{Fz}\right\rangle .$ In
the opposite, dirty (diffusive or Maxwell) limit, $P_A$ is
determined by $P_{Nv^{2}}\propto
  \left\langle N(E_{F})v_{Fz}^{2}\right\rangle $%
\cite{6,M}.
The same $P=P_{Nv^{2}}$ characterizes the spin polarization of the bulk
electric current, as well as the tunneling current
in the case of specular, low transparency barrier \cite{M}.
In the Ni$_{x}$Fe$_{1-x}$ system, where
the transport properties are determined by both heavy d-electrons and light
s-electrons, the tunneling current as well as the current in the
diffusive case of Andreev reflection  are dominated by the majority spins, even
though their density of states is smaller.
Similarly, there is no
reason to believe that  $P_T$, or $P_A$ should be related to the 
magnetic moment,
which is defined as a difference in the total number of spin-up and spin-down
electrons. Since in Ni$_{x}$Fe$_{1-x}$ the Fermi surface is far from spherical,
the effective mass is strongly dependent on the wave vector, and the bands are
highly hybridized, it is unrealistic to expect that the 
spin-dependent transport  of
these compounds can be described by the simple model of the polarized 
homogeneous
electron gas, as it was often assumed in earlier works. Consequently, the
once popular idea that the spin polarization, as measured by the tunneling
spectroscopy, is proportional to the bulk magnetization
\cite{4,5} is not applicable to this system.

In this article, we present direct detailed measurements of the transport
spin-polarization of the Ni$_{x}$Fe$_{1-x}$
system by a newly developed Point Contact Andreev Reflection (PCAR) 
technique \cite{6}.
We also perform band structure calculations of the spin polarization
in this system,
using a standard
LSDA technique. The measured values of the
transport spin polarization are almost independent of the composition
\cite{agree}, in reasonably good agreement with the theory. Based
on the band structure calculations, we interpret this surprising result as
a consequence of compensation of the numerous but heavy d-electrons and
scarce but light s-electrons.

Many thin films and bulk samples
were studied. They included a Ni single crystal, several Ni and Fe
polycrystalline foils,
a [100]--oriented single
crystal Fe film grown on a GaAs substrate by molecular beam epitaxy, and a
number of variable composition Ni$_{x}$Fe$_{1-x}$ films grown on
Si-[100] substrates deposited by thermal (e-beam) evaporation \cite{9}.
In order to  make meaningful conclusions from the
measurements and to compare the experimental results with the theory
  we determined the structural phase of the Ni$_{x}$Fe$
_{1-x}$ films for the entire composition range. 
X-ray diffraction data
(specular $\theta /2\theta $-scans) were recorded for each of
the Ni$_{x}$Fe$%
_{1-x}$ compositions over
two angular ranges, 35$^{o}$-68$^{o}$ and 71$^{o}$%
-86$^{o}$. In all cases only a single phase
was found \cite{11}%
: the $\gamma $(FCC)-phase is present for $x>0.47$; the $\alpha $%
(BCC)--phase is present for $x<0.30$. These results are consistent with the
results for bulk samples \cite{12} and for thin films \cite{13}.
The lattice
parameters for the films were within 0.3\% of the
corresponding bulk values.

The details of the PCAR technique are
described elsewhere \cite{6}. The method measures
the degree of suppression of Andreev
reflection at a ferromagnet/superconductor interface due to
the spin polarization of the ferromagnet\cite{15}. The Andreev
process allows propagation of a single electron with
the energy below the superconducting gap $\Delta $ from the normal metal to the
superconductor, by reflecting at the
interface as a hole via a time reversal
process. In a non-magnetic normal metal this is always allowed,
because in such a metal each energy state has both
spin-up and spin-down electrons. However, in a magnetic metal this is no
longer true and Andreev reflection is limited by the spin direction
with the smaller number of conductance channels,
which drastically changes the sub-gap conductance.
To account for finite temperatures and arbitrary
barrier transparancy,
$Z$ the  normalized conductance data $G(V)/G_{n}$
  ($G_n$ is obtained at
voltages $V\gg \Delta/e$,$e$ is the electron charge) were compared to the
modified \cite{6} Blonder, Tinkham and Klapwijk (BTK)\cite{16}
model with only two adjustable parameters ($P$ and $Z$)\cite{17}. The
temperature was generally taken to be equal to the temperature of the helium
bath and $\Delta $ was defined separately from the BCS dependence
\cite{Dynes}.
This procedure allowed us to determine the {\it magnitude} of the spin
polarization \cite{6}.

\vspace{.2in} \begin{figure}[tbp]
\centerline{ \epsfig{file=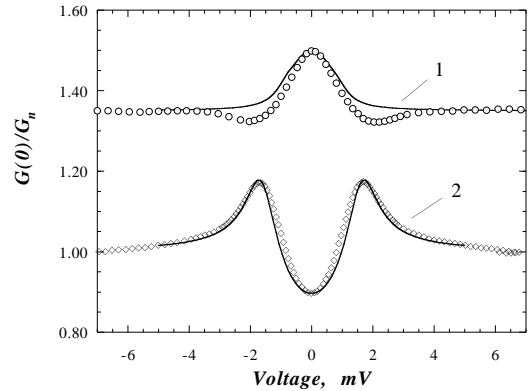,width=0.95\linewidth} }
\setlength{\columnwidth}{3.2in} \nopagebreak \caption{
$G(V)/G_n$ as a function of voltage $V$ for two samples:
1(o)- Fe point/Ta foil base; 2($\diamondsuit $) 
- Nb point/Ni$_{80}$Fe$_{20}$ film base;
solid curves -- modified BTK fits ($T= 1.7$ K, $Z= 0$, $
\Delta = 0.6$ meV, and $T=2.2$ K, $Z=0.4$, $\Delta=1.5$ meV, respectively).
The values of {\it P} obtained from these fits are 42.5\% for Fe
and 49.5\% for Ni$_{80}$Fe$_{20}$. Note that the bath temperature in the
latter case was $1.8$ K (see text). 
Curve 1 has an arbitrary vertical offset.
}
\end{figure}

Our adjustment mechanism consisted of a sharpened rod
(superconducting or ferromagnetic) which was driven by a micrometer until it
touched the (ferromagnetic or superconducting) base. Superconducting Nb, V,
and Ta were used for the measurements reported here. Typical normalized
conductance data $G(V)/G_{n}$ obtained by the PCAR method are shown in Fig.1
as a function of voltage {\it V}. For each sample a number of
different contact junctions (with the contact resistance 1 $\Omega <R_{c}<$%
100 $\Omega $) were measured and then fitted with the modified BTK model. In
Table I we present a summary of the data obtained for several samples 
for the end points (Ni and Fe) which were studied in more detail. Although we
observed some variation in the values of
{\it P }for the same material, the results are quite consistent and do not
appear to depend strongly on whether the ferromagnet was a bulk single
crystal, a foil or a film. Furthermore, it does not seem to matter whether
it was the point or base in the contact. Finally, the value of {\it P} does
not depend strongly on the superconducting material. Accordingly, the
values for $P$ for individual samples of each material were averaged together.
For Fe, $\left\langle P\right\rangle = (44 \pm  3)$\% and for Ni,
$\left\langle P\right\rangle = (46 \pm  3)$\%.

The PCAR results for the entire Ni$_{x}$Fe$_{1-x}$
are shown in Fig. 2. For the measurements of the thin film series
a Nb tip was used.
The spin polarization is almost
composition-independent, whereas the measured magnetic moment (shown in the
inset in Fig.2) changes by a factor of three.
Evidently, our
measurements do not show any correlation between the spin
polarization and magnetic moment, which was observed in the early tunneling
spectroscopy measurement\cite{18}. Although our spin
polarization values differ substantially from these early results, they are
quite close to the most recent tunneling measurements\cite{9}
  obtained from the
``companion'' Ni$_{x}$Fe$_{1-x}$ samples \cite{agree}.
This result is not necessarily to be expected as PCAR probes
$N(E_{F})v_{Fz}^{2}$
averaged over the entire Fermi surface, whereas tunneling through a
thick barrier can be shown to probe $N(E_{F})v_{Fz}^{2}$ only at those
selected points of the Fermi surface where quasi-momentum is
perpendicular to the interface. Apparently, averaging over individual
grains in the Ni$_{x}$Fe$_{1-x}$ films helps to bring the tunneling
spin-polarization results close to the Fermi surface-averaged PCAR
results.

\vspace{-.0in}
\begin{figure}[tbp]
\centerline{\epsfig{file=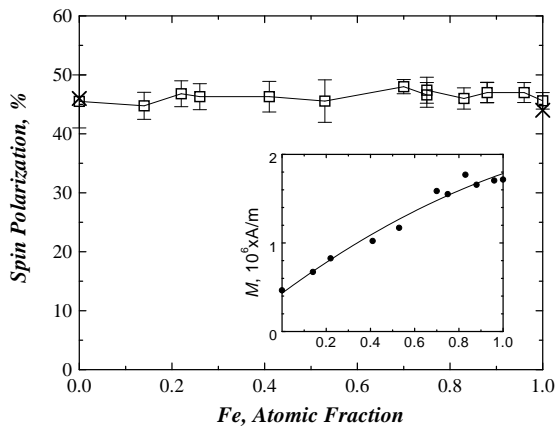,width=.95\linewidth}}
\vspace{.1in} \setlength{\columnwidth}{3.2in}
\caption{PCAR results for the 
spin polarization {\it P} as a function of Fe content for Ni$%
_{x} $Fe$_{1-x}$
samples. ($\square $) - films;
($\times $) - pure Ni and Fe foils and single crystals.
Inset: Magnetic moment, {\it M}, as a function of the
Fe content for Ni$_{x}$Fe$_{1-x}$ samples. Lines are guides to the
eye.}
\end{figure}

To calculate the spin polarization, we performed
LSDA band structure calculations\cite{20}. Our X-ray measurements
allowed us to conclude that a single structural phase was present at any
given Ni-Fe composition. Thus we were able to compare the experimental
results with the calculations performed in the appropriate lattice
structure. For Ni content $x<0.35$, the calculations were carried out in an
average BCC lattice, for $x>0.35$ an average FCC lattice was used\cite{21}.
Several ordered Ni-Fe
supercells with the compositions NiFe$_{15}$, NiFe$_{7}$%
, NiFe$_{3}$, NiFe$_{2}$, Ni$_{3}$Fe, and Ni$_{7}$Fe were
used. The results of the calculations of the spin polarization ($%
P_{N},P_{Nv}$ and $P_{Nv^{2}}$) are shown in Fig. 3. 
First of all, we observe that the three polarizations
are dramatically different, which emphasizes once again the importance of
the correct definition of the spin polarization for a given experiment.
These differences are due to the  strong variation of the
kinematic properties between s-like and d-like electrons. Specifically, the
Fermi velocity anisotropy between the different sheets of the Fermi surface,
as well as the angular anisotropy, 
have to be taken
into account for a
quantitative description of any spin-transport experiment.
While ``heavy'' parts of the Fermi surface dominate the DOS spin
polarization, ``light parts'' determine the spin polarization relevant for
transport and tunneling phenomena\cite{M}. 
There is good agreement between the experimental data for the Ni-rich
and Ni-poor alloys and band structure calculations for the diffusive limit, $%
P_{Nv^{2}}$ (except for pure Fe where $P_{Nv}$ agrees with the
experiment better than $P_{Nv^{2}}$). This result is quite reasonable
because the electron mean free
path, $l_{e}$, of these alloys  (but not necessarily for 
the pure components) is typically very short
(compared to the size of the contact) even at low
temperatures, especially for minority spins ($l_{e}\sim$ 5-10 \AA )
\cite{23}.
This is also consistent with the agreement between
PCAR and tunneling spin polarizations\cite{9}; as
mentioned above, the latter is also
  defined by $\langle Nv^2\rangle$.

\vspace{-.2in}
\begin{figure}[tbp]
\centerline{\epsfig{file=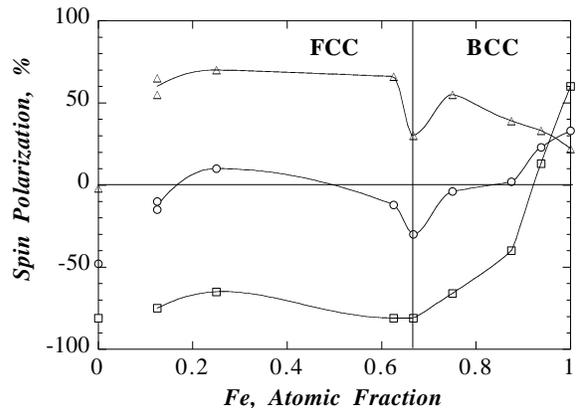,width=1.3\linewidth}}
\vspace{-.5in} \setlength{\columnwidth}{3.2in}
\caption{
The results of the band structure calculations for $P_{N},P_{Nv}$ and $%
P_{Nv^{2}}$: ($\square $) - {\it P}$_{N}$; (o) - {\it P}$_{Nv}$; ($\triangle $%
) - {\it P}$_{Nv^2}$. Lines are guides to the eye. The results for pure
Ni are shown for completeness.} \end{figure}
We could not perform reliable calculations for pure Ni.
This reflects a well-known problem in conventional band
structure theory, which is unable to completely account for  electronic
correlation effects in the 3d-states of metallic oxides and, to a
lesser extent, of Ni. The
correlation effects in Ni
are known to reduce the exchange splitting by a
factor of two which, in turn, should affect the spin
polarization.
For a different reason, we exclude the theoretical calculations for the
compounds close to the 50:50 Ni:Fe composition. At relevant lattice
parameters the FCC phase of Fe is
antiferromagnetic, so close to its solubility limit in the FCC Ni
(approximately 60-65\%) the corresponding Ni-Fe alloys must have Fe clusters
which are sufficiently large to develop antiferromagnetic order. On the
other hand, Fe-Ni and Ni-Ni interactions are ferromagnetic. This creates
frustration leading to non-collinear spin ordering\cite{2}. A theory of
spin-polarized transport in such systems is yet to be developed.

In summary, we have presented the band structure calculations of the
transport spin polarizations in the Ni-Fe system and the experimental
measurements of the same system using the PCAR technique. Overall, the spin
polarization measured by PCAR technique agrees reasonably well with
the band structure calculations for $%
P=P_{Nv^{2}}$. It is also in surprisingly good agreement with
the most recent tunneling results \cite{9}. Furthermore, our results
repudiate the
idea of a direct relationship between the spin polarization and the
magnetic moment. At the same time we show that
the spin polarization in electronic transport is determined by
the delicate balance of the density of states and the kinematics of the s-
and d- electrons
(the variation of the Fermi velocity over the Fermi surface) and, therefore,
dependent on the measurement technique and the transport process in
question. In particular, our calculations give a quantitative explanation
for a long-standing problem of the positive values of tunneling spin 
polarization
observed for the Ni-Fe system, which has important implications both for 
fundamental
issues of spin transport and for magnetoelectronics applications.

We are grateful to J. S. Moodera for providing the results on
tunneling\cite{9} prior to publication, G. Prinz for useful
discussions,
and  T. Ambrose, C. T. Tanaka, T.J.M.  Verspaget, and
M. Maoliyakefu for technical assistance.
We also thank J. Byers for
numerous discussions and providing the modified
BTK program, and P. Broussard for help in computer modeling.
This work was supported by ASEE and ONR.


\begin{table}[tbp]
\caption{Spin polarization results for pure Fe and Ni. $N$ refers
to the number of distinct point contacts made;
{\it P} to the average polarization obtained for the $N$
contacts, and $dP$ to the standard deviation.
C=crystal, Fl=foil, Fm=film.}
\begin{tabular}{lcc|lcc}
Point/Base & $N$ & {\it P}(\%)&
Point/Base & $N$ & {\it P}(\%)\\
\tableline Fe/V (C) & 9 & 45$\pm$2 &
Nb/Ni (C-1) & 8 & 45$\pm$2 \\
Fe/Ta (Fl) & 14 & 46$\pm$2 &
Nb/Ni (C-2) & 8 & 41$\pm$4 \\
Fe/Nb (Fl) & 3 & 42$\pm$3 &
Nb/Ni (C-3) & 11 & 48$\pm$4 \\
Nb/Fe (Fm) & 5 & 41$\pm$3 &
Nb/Ni (Fl) & 10 & 45$\pm$2 \\
Ta/Fe (Fm) & 12 & 45$\pm$2 &
Nb/Ni (Fm) & 14 & 45$\pm$3 \\
&&&Ta/Ni (Fl-1) & 8 & 44$\pm$4 \\
&&&Ta/Ni (Fl-2) & 10 & 50$\pm$1 \\
\tableline Average, Fe &$\left\langle P\right\rangle $   & 44$\pm$ 3 &
  Average, Ni & $\left\langle P\right\rangle $ & 46$\pm$ 3
\end{tabular}
\end{table}

\begin{references}
\vspace{-.7in}
\bibitem{1}  G.A. Prinz, Physics Today {\bf 48,} 58 (1995); G.A. Prinz,
Science {\bf 282}, 1660 (1998).

\bibitem{3}  H. Ebert, P. Strange, and B.L. Gyorffy, J. Phys. F {\bf 18},
L135 (1988), and reference therein.

\bibitem{4} J.B. Gadzuk, Phys. Rev. {\bf 182}, 416
(1969); M.B. Stearns, J. Magn. Mag. Mater. {\bf 5}, 167 (1977).

\bibitem{9} J.S. Moodera {\it et al},
in preparation.

\bibitem{6}  R. J. Soulen et al, Science {\bf 282}, 85 (1998).

\bibitem{M}  I. I. Mazin, Phys. Rev. Lett.,  {\bf  83}, 1427 (1999).

\bibitem{B}S.K. Upadhyay {\it et al},
Phys. Rev. Lett. {\bf 81}, 3247 (1998).

\bibitem{5}  R. Meservey, D. Paraskevopoulos and P.M. Tedrow, Phys. Rev.
Lett. {\bf 37}, 858 (1976).

\bibitem{agree} ``Companion'' samples
of the same composition made under the same conditions (see
R. van der Veerdonk, Ph.D. Thesis,
Eindhoven University, Eindhoven, 1999) and were
also measured by Moodera {\it et al}
\cite{9} using standard tunneling spectroscopy \cite{10}.

\bibitem{11}  Based on the signal-to-noise ratios of the strongest peak in
our diffraction patterns, it is estimated that a second phase could be
detected if present in quantities greater than 5\%.

\bibitem{12}  A.J. Bradley and A. Taylor, Phil. Mag. {\bf 23}, 545 (1937).

\bibitem{13}  M. Konno and H. Konno, J. Magn. Mag. Mater. {\bf 118}, 381
(1993).

\bibitem{15}  M.J.M. de Jong and C.W.J. Beenakker, Phys. Rev. Lett.
{\bf 74}, 1657 (1995).

\bibitem{16}  G.E. Blonder, M. Tinkham and T.M. Klapwijk, Phys Rev. B. {\bf %
25}, 4515 (1982).

\bibitem{17}  Note that our analysis 
is based on an assumption (used both in the
BTK paper\cite{16} and in our modified version\cite{6})
of a ballistic contact,
for which the mean free path $l_{e}$ is
much larger than the size of the contact, $d$.
It is difficult to assess the effect of violation
 of 
the condition ($ l_{e}\gg d$). Preliminary
calculations show, however, that it mainly affects
the values of  $Z$ in our fits, while the changes
in the resulting spin polarization are relatively small.

\bibitem{Dynes} For some junctions we could obtain
better fits by using a higher temperature.
This is most likely due to the fact that we did not explicitly introduce
any gap smearing, as it is customary in superconducting tunneling
(i.e. R.C. Dynes,  Phys. Rev. Lett. {\bf 53} 2437 (1984).
Fitting with an increased temperature is to a large extent equivalent to
an additional gap smearing. We emphasize that even when
increasing the temperature
parameter impoves the quality of the fits, it hardly changes the 
resulting  spin polarization.

\bibitem{18}  D. Paraskevopoulos, R. Meservey, and
P. M Tedrow, Phys. Rev. B {\bf 16}, 4907 (1977).
In these measurements, a two-step junction fabrication
process was used, while the more recent experiments\cite{9} used
{\it in situ } fabrication. Tunneling spectroscopy is
a more surface sensitive technique \cite{10}
than PCAR and, therefore, may be susceptible to surface conditions
of nickel, which are clearly improved by using
an {\it in-situ} process\cite{9}.

\bibitem{20}  The LMTO-ASA Stuttgart code was used, see e.g. O.K. Andersen
and O. Jepsen, Phys. Rev. Lett. {\bf 53}, 2571 (1984).

\bibitem{21}  Using the correct structural phase in these calculations is
important (replacing fcc by bcc and vice versa yields significantly
different results).

\bibitem{23}  D.Y. Petrovykh et al, Appl. Phys. Lett.{\bf \ 73,} 3459 (1998).
\bibitem{2}  Y. Wang et al, J. Appl. Phys. {\bf 8}, 3873 (1997).

\bibitem{10}  P.M. Tedrow and R. Meservey, Phys. Rep. {\bf 238}, 173 (1994).
\end{references}
\end{document}